\documentclass{aa}

\usepackage{txfonts}
\usepackage{graphicx}

\newcommand{\grs}     {GRS 1758$-$258}
\newcommand{\unoe}    {1E 1740.7$-$2942}
\newcommand{\gr}      {^{\circ}}
\newcommand{\pr}      {^{\prime}}
\newcommand{\prpr}    {^{\prime \prime}}
\newcommand{\grp}     {{\rlap.}^{\circ}}
\newcommand{\prprp}   {{\rlap.}^{\prime \prime}}
\newcommand{\segp}    {{\rlap.}^{s}}

\begin{document}

\title{Identification of the optical and near-infrared counterpart of \grs}

\author{
A.~J. Mu\~noz-Arjonilla\inst{1,2}
\and J. Mart\'{\i}\inst{1,2}
\and P.~L. Luque-Escamilla\inst{3,2}
\and J.~R. S\'anchez-Sutil\inst{2}
\and E. S\'anchez-Ayaso\inst{1,2}
\and J.~A. Combi\inst{4}
\and I.~F. Mirabel\inst{5}
}

\offprints{A.~J. Mu\~noz-Arjonilla}

\institute{
Departamento de F\'{\i}sica, EPS,  
Universidad de Ja\'en, Campus Las Lagunillas s/n, Edif. A3, 23071 Ja\'en, Spain \\
\email{ajmunoz@ujaen.es, jmarti@ujaen.es, esayaso@ujaen.es}
\and
Grupo de Investigaci\'on FQM-322, 
Universidad de Ja\'en, Campus Las Lagunillas s/n, Edif. A3, 23071 Ja\'en, Spain \\
\email{jrssutil@ujaen.es}
\and
Dpto. de Ing. Mec\'anica y Minera, EPS,  
Universidad de Ja\'en, Campus Las Lagunillas s/n, Edif. A3, 23071 Ja\'en, Spain \\
\email{peter@ujaen.es}
\and
Instituto Argentino de Radioastronom\'{\i}a (CCT La Plata, CONICET), 
C.C.5, (1894) Villa Elisa, Buenos Aires, Argentina \\
\email{jcombi@fcaglp.unlp.edu.ar}
\and
Laboratoire AIM, IRFU/Service d'Astrophysique, Bat. 709, CEA-Saclay, 91191 Gif-sur-Yvette Cedex, 
France \& \newline
Instituto de Astronom\'{\i}a y F\'{\i}sica del Espacio (CONICET--UBA), CC 67, Suc. 28, 1428 Buenos 
Aires, Argentina \\
\email{felix.mirabel@cea.fr}
}

\date{Received / Accepted}

\titlerunning{Identification of the optical and near-infrared counterpart of \grs}

\abstract
{Understood to be a microquasar in the Galactic center region, \grs\ has not yet been unambiguously 
identified to have an optical/near-infrared counterpart, mainly because of the high absorption and 
the historic lack of suitable astrometric stars, which led to the use of secondary astrometric 
solutions. Although it is considered with \unoe\ as the prototypical microquasar in the Galactic 
center region, the Galactic origin of both sources has not yet been confirmed.}
{We attempt to improve previous astrometry to identify a candidate counterpart to \grs. We present 
observations with the Gran Telescopio de Canarias (GTC), in which we try to detect any powerful 
emission lines that would infer an extragalactic origin of this source.}
{We use modern star catalogues to reanalyze archival images of the \grs\ field in the optical and 
near-infrared wavelengths, and compute a new astrometric solution. We also reanalyzed archival 
radio data of \grs\ to determine a new and more accurate radio position.}
{Our improved astrometric solution for the \grs\ field represents a significant advancement on 
previous works and allows us to identify a single optical/near-infrared source, which we propose as 
the counterpart of \grs. The GTC spectrum of this source is however of low signal-to-noise ratio 
and does not rule out a Galactic origin. Hence, new spectral observations are required to confirm 
or discard a Galactic nature.}


\maketitle

\section {Introduction}

\grs\ is one of the two brightest persistent hard X-ray sources in the Galactic center region and 
belongs to the class of Galactic microquasars (see e.\,g., \cite{mirabel-99} for a general 
discussion of microquasar properties). Originally discovered in hard X-rays (\cite{sunyaev-91}; 
\cite{goldwurm-94}), \grs\ was classified as a microquasar after the detection of radio bipolar 
jets emanating from it, extending over $\sim$\,1$\pr$ and ending with the formation of two radio 
lobes (\cite{rodriguez-92}). Although it is considered, together with \unoe\ (\cite{mirabel-92}), 
the prototypical microquasar in the Galactic center region, its morphology is reminiscent of a 
Fanaroff-Riley II galaxy and an extragalactic origin cannot be ruled out.

The radio source proposed as the exciting core of the \grs\ microquasar lobes is found to coincide 
with a {\it Chandra} X-ray source (\cite{heindl-02}; \cite{marti-02}). These X-ray and radio 
observations provide us with a very accurate sub-arcsec position, which is essential to find an 
optical/near-infrared counterpart in the crowded fields of the Galactic center region. 
Unfortunately, efforts made in this sense have been unable to unambiguously propose a single source 
as the possible counterpart to \grs, mainly due to the absorption towards the Galactic center 
region and the early lack of suitable astrometric standards (\cite{marti-98}; \cite{rothstein-02}). 

In this work, we have taken advantage of the existence of modern star catalogues, which allow us to 
improve the astrometry of the \grs\ field in the optical and near-infrared wavelengths. We also 
present a new and more accurate radio position of this microquasar that we derived based on a 
revision of archival data. As a result, we find that there is only one optical/infrared object 
astrometrically coincident with the accurate \grs\ X-ray and radio positions, which we propose as 
its counterpart. Furthermore, we report the first results of a follow-up observation with the Gran 
Telescopio de Canarias (GTC).

\section {\grs\ revisited in the radio}

\begin{figure*}[htpb]
\begin{center}
\includegraphics[scale=0.4]{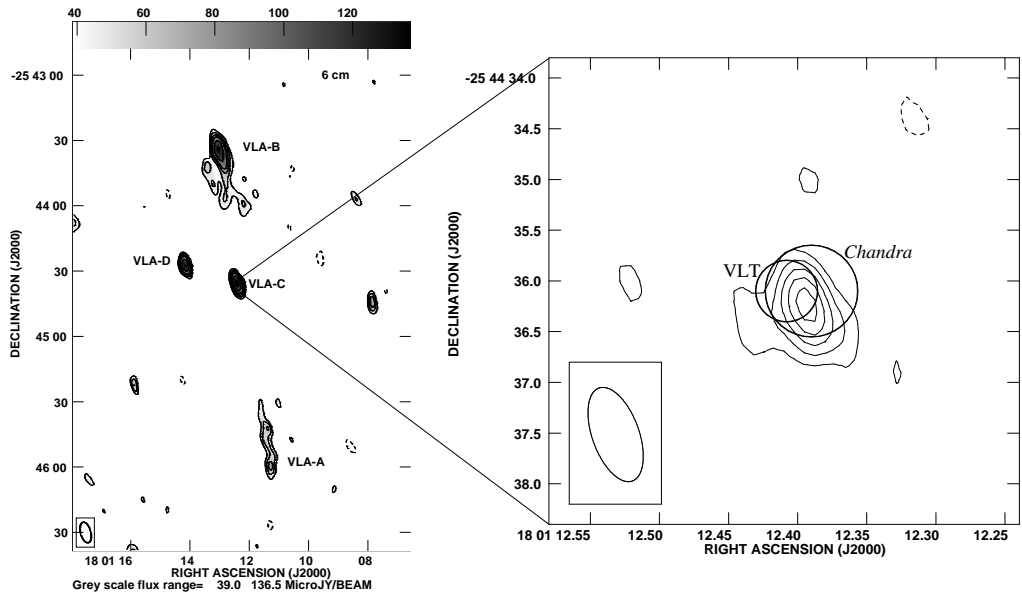}
\caption{{\bf Left.} Large-scale radio map of \grs\ at 6~cm wavelengths (adapted from 
\cite{marti-02}). The central core of the microquasar is labelled VLA-C. The collimated jets end 
with the formation of two radio lobes (VLA-A and VLA-B). VLA-D is considered to be a likely 
non-related radio source. {\bf Right.} Central core of \grs\ at 6~cm, obtained in this work after 
combining all the A-configuration VLA data in Table~\ref{table-VLA-obs}. Radio contours correspond 
to -3, 3, 5, 7, 9, and 11 times the rms noise of 30~$\mu$Jy. The synthesized beam of this high 
resolution map is shown at the bottom left corner, its size being $0\prprp 97 \times 0\prprp 46$ 
with a position angle of $19\grp 53$. The 90\% confidence error circles of {\it Chandra} and VLT 
positions are also shown.}
\label{figura1}
\end{center} 
\end{figure*}

\begin{table}
\begin{center}
\caption[]{VLA archival observations used in this paper \label{table-VLA-obs}}
\begin{tabular}{cccc}
\hline
\hline
Date  &  Project & \# of visibilities & Time on source \\
\hline 
2003 Jul 31 & AR523 & 249953  & 2807 s \\
2004 Oct 16 & AR545 & 200182  & 1280 s \\
2006 Jan 30 & AR570 & 151035  & 2153 s \\
\hline
\hline \end{tabular}
\end{center}
\end{table}

The National Radio Astronomy Observatory (NRAO) archive contains huge amounts of radio data from 
different sources that has accumulated over the years. Sometimes, several observations can be 
combined to produce a single deep radio map and reanalyzed with an aim different from the one it 
was originally planned. Similar approaches have been conducted by our group for X-ray binaries 
showing how useful archival data can be (see e.\,g. \cite{sanchez-08}; \cite{munoz-09}).

In the context of this work, we explored the NRAO archive to improve the position of the core of 
\grs. For this purpose, we restricted the selection criteria to those experiments carried out with 
the Very Large Array (VLA) in its extended A-configuration and with on-source times higher than a 
thousand seconds. Table~\ref{table-VLA-obs} shows the observations used in this paper, all of them 
at the 6~cm wavelength. The total on-source time amounted to almost 2~h. Observations were made in 
two IF pairs of 50~MHz bandwidth each. The phase calibrator used was always 1751-253 which is 
located $2\gr$ away from the target and has a position code B in the VLA Calibration 
Manual\footnote{$<$http://www.vla.nrao.edu/astro/calib/manual/csource.html$>$}. 
The data of each project were separately processed using the AIPS software package of NRAO 
following the standard procedures for continuum calibration of interferometers. All data sets were 
merged into a single {\it uv} file, from which we computed our final map using natural weights. 
Radio contours corresponding to the core of \grs\ are shown in the right panel of 
Fig.~\ref{figura1}. Our new and more accurate radio position, namely 
R.A.(J2000)$= 18^h 01^m 12\segp40$ and DEC.(J2000)$=-25\gr 44\pr 36\prprp3$, was fitted in the 
image plane. Although it has a statistical error of $0\prprp 04$ in both coordinates, we 
conservatively adopt an uncertainty of $0\prprp 1$ to consider any possible systematics. Our 
declination value differs $0\prprp5$ from a previous result also derived with the VLA at 6~cm in 
the A-configuration and using the same calibrator source (\cite{mirabel-94}). This apparent 
discrepancy may be caused by the position of the phase calibrator having been refined since those 
old observations, leading to a shift of nearly $0\prprp3$ almost completely in declination towards 
the south. Hence, our result represents an improvement of the radio position of the \grs\ core with 
respect to previous measurements (\cite{mirabel-94}; \cite{marti-02}).

\section {The optical/near-infrared counterpart of \grs}

\begin{figure*}[htpb]
\begin{center}
\includegraphics[scale=0.5]{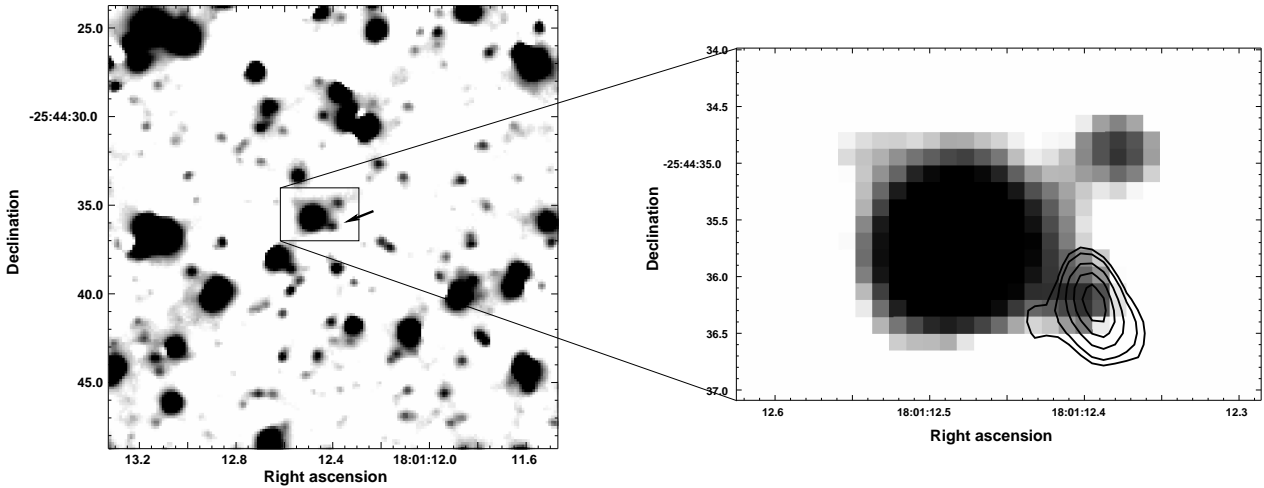}
\caption{{\bf Left.} $K_s$--band image corresponding to the central core of \grs\ taken with the VLT 
and the ISAAC instrument in 2003. The arrow points at the location of the near-infrared candidate 
counterpart. {\bf Right.} Zoomed detail of the map on the left. Radio contours corresponding to the 
central core of \grs\ at 6~cm are shown, superimposed on the $K_s$--band image. Contour levels are 
4, 5, 7, 9, and 11 times the rms noise of 30~$\mu$Jy.}
\label{figura2}
\end{center} 
\end{figure*}

\subsection{Reanalyzing archival data}

The improved radio position of \grs, and the present day availability of modern star catalogues in 
the near-infrared, such as the Two Micron All Sky Survey (2MASS), motivated us to refine the old 
astrometry, which had been obtained in two steps because of the historic lack of suitable reference 
stars in the field (\cite{marti-98}). The total combined error in this old paper was close to 
$1\prpr$, and a candidate counterpart could not be unambiguously proposed at that time.

Hence, we reanalyzed archival images at optical and near-infrared wavelengths of the \grs\ field. 
In particular, we used images in the $R$ and $I$ bands obtained with the ESO New Technology 
Telescope (NTT) in 1998, and in the $K_s$ band taken with the Very Large Telescope (VLT) in 2003 in 
a project by Heindl and collaborators. Astrometry was recalculated based on the data of more than 
30 stars within our field whose accurate positions were retrieved from the 2MASS catalogue. The 
total combined astrometric error is $0\prprp 16$ for the NTT maps and $ 0\prprp 18$ for the VLT 
image. As a result, among the three candidate counterparts of \grs\ that were proposed almost a 
decade ago, only one of them remains consistent with both the VLA and {\it Chandra} positions (see 
Figs.~\ref{figura1} and \ref{figura2} and their right panel zooms). Approximate photometry was 
calculated in the $K_s$ image by estimating the zero point after comparing the instrumental 
magnitudes of ten stars in the field with the corresponding ones in the 2MASS catalogue. The $K_s$ 
magnitude derived here for \grs\ is clearly consistent with previous measurements 
(\cite{eikenberry-01}). Magnitudes in the $R$ and $I$ bands were directly retrieved from the 
original work (\cite{marti-98}). These apparent magnitudes are listed in Table 
\ref{table-magnitudes}.

\subsection {GTC follow-up observations}

We carried out spectroscopic observations using the 10~m GTC and the OSIRIS 
instrument\footnote{$<$http://www.iac.es/project/OSIRIS/$>$} at the Observatorio de La Palma 
(Spain) in long-slit mode. Our goal was to obtain a spectrum of the optical candidate counterpart 
of \grs, and the GTC with OSIRIS in a low-resolution spectral mode seemed an appropriate 
combination to use for such a weak source. The idea was to identify any absorption or emission 
feature that could allow us to discriminate between a Galactic or an extragalactic origin for this 
counterpart of \grs, and therefore to know if we should carry on considering it as an archetypical 
Galactic microquasar.

\begin{figure}[htpb]
\begin{center}
\includegraphics[scale=0.3]{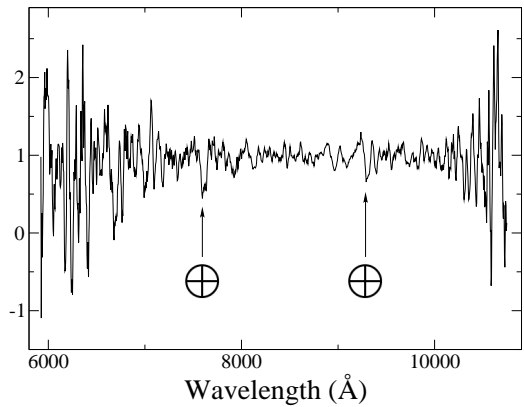}
\caption{Noisy spectrum of the proposed \grs\ optical counterpart. Only telluric features ($\oplus$ in the figure) are obvious.}
\label{figura3}
\end{center} 
\end{figure}

The observations were performed on May 21 and June 30 2009. The slit was aligned with the 
parallactic angle to minimize light loss due to atmospheric refraction. The grism covered the 
3700 -- 10\,000 $\AA$ spectral range, and the integration time amounted to 2.3 hours. All frames 
were reduced using standard procedures for sky background subtraction, flat-fielding, and 
extraction of the spectrum, based on the IRAF\footnote{$<$http://iraf.noao.edu/$>$} image 
processing system. Unfortunately, the seeing conditions, the weakness of our target, and our high 
air-mass value even at transit time rendered the observation very difficult. Figure~\ref{figura3} 
shows the resulting noisy spectrum, after wavelength calibration and continuum normalization. The 
shortest wavelengths covered by the grism are not shown since no signal is detected above the 
noise.

\section {Discussion}

Figure~\ref{figura2} again shows the \grs\ radio map superimposed on the VLT $K_s$--band image. The 
proposed optical/near-infrared candidate counterpart, which is the only one that is astrometrically 
coincident with {\it Chandra} and VLA positions of the core of \grs, is clearly seen. It turns out 
to be a weak and highly absorbed source that is barely detected towards the blue. The offsets from 
the improved radio position are $\Delta \alpha \cos \delta \simeq 0\prprp 16$ and 
$\Delta \delta \simeq 0\prprp 14$, which are clearly consistent with astrometric errors.

The GTC spectrum of the candidate counterpart of \grs\ shown in Fig.~\ref{figura3} is probably also 
contaminated by those of the two stars that are just $\sim 1\prpr$ away. Apart from some telluric 
features, no spectral lines were confidently observed. Nevertheless, the lack of strong emission 
lines, which would be visible even in a contaminated spectrum if \grs\ were a sort of nearby active 
galaxy (e.\,g., a Seyfert or a FR-II), is consistent with the usual interpretation of \grs\ as a 
source within the Galaxy. However, in this case, the eccentricity of the orbit of the star would 
need to be very high to fill its Roche lobe if a 18.45~d period were assumed (\cite{rothstein-02}; 
\cite{smith-02}). The problems raised by this inconsistency will require further attention to 
unveil the true physical scenario behind \grs.

In the absence of conclusive spectroscopic data, only broad-band photometry can place broad 
constraints on the physical nature of \grs. An interstellar extinction of $A_V \simeq 8.4$ mag was 
estimated following a similar approach to that of \cite{predehl-95} and considering a column 
density towards \grs\ of $N_H \simeq (1.5 \pm 0.1) \times 10^{22}$ cm$^{-2}$ 
(\cite{mereghetti-97}). We also computed the values $A_K$, $A_I$, and $A_R$ using the relations 
reported by \cite{rieke-85}. Assuming a Galactic center distance of 8.5~kpc, dereddened magnitudes 
of the candidate counterpart to \grs\ are obtained. The results of our final revised photometry are 
summarized in Table~\ref{table-magnitudes}. We searched for any possible variability of the 
proposed counterpart in the ESO archives. A total of 11 observing nights of data were available, 
obtained with VLT and the ISAAC instrument in the $K_s$--band. No variability was detected with 
amplitude larger than $\pm 0.5$~mag on a timescale of weeks. The $K_s$ dereddened magnitude is 
roughly consistent with an early A-type main sequence star. However, the corresponding colours 
$(R-I) \simeq -0.8$ and $(I-K) \simeq +1.2$ are inconsistent with this or any other spectral type. 
In a Galactic context, this may be indicative of the optical/near-infrared luminosity being 
dominated by a non-stellar component (e.\,g., an accretion disk) as we would expect from a low-mass 
X-ray binary. On the other hand, we cannot rule out the possibility that this might be caused by 
the brighter star ($K_s \simeq 13.7$), which is located 
very close to our proposed candidate counterpart and may contaminate the photometry even after we 
have carefully tried to subtract its effects. Since a different extinction law may explain the 
discrepant colours, we computed them again following the conversions of near-infrared extinctions 
to $A_V$ towards the nuclear bulge described by \cite{gosling-09}. Despite the different form of 
the extinction law, the derived colours do not substantially change compared to those listed in 
Table~\ref{table-magnitudes}. Hence, our conclusion of contamination by either the nearby star or 
by a non-stellar component such as an accretion disk remains unaltered.

\begin{table}
\begin{center}
\caption{\label{table-magnitudes} Magnitudes of the candidate counterpart to \grs}
\begin{tabular}{ccccc}
\hline
\hline
Filter & Observation &            Apparent          &   Interstellar  &  Dereddened   \\ 
       & date        &            magnitude         &   extinction    &  magnitude    \\ 
\hline
$R$    & 1998 Mar 26 & $22.6 \pm 0.3$ $^{\rm{(a)}}$ & $6.3  \pm 0.5$  & $1.7 \pm 0.6$ \\ 
$I$    & 1998 Mar 26 & $21.1 \pm 0.3$ $^{\rm{(a)}}$ & $4.0  \pm 0.3$  & $2.5 \pm 0.5$ \\ 
$K_s$  & 2003 Aug 23 & $16.9 \pm 0.2$ $^{\rm{(b)}}$ & $0.94 \pm 0.07$ & $1.3 \pm 0.3$ \\ 
\hline
\hline
\end{tabular}
\end{center}
$^{\rm (a)}$ Magnitude retrieved from a previous work (\cite{marti-98}) \\
$^{\rm (b)}$ Magnitude computed in the present work \\
\end{table}

Despite the challenge to interpret spectroscopic observations, the main contribution of this work 
has been the identification of a serious candidate to the optical/near-infrared counterpart of 
\grs\ based on accurate astrometric coincidence. This opens the possibility of investigating the 
true nature of this source. Additional spectroscopic observations with 8~m-class telescopes and 
adaptive optics at southern locations are required to confirm its Galactic (or extragalactic) 
origin and, in the former case, to more clearly classify its spectral type.

\begin{acknowledgements}
{\small The authors acknowledge support by grant AYA2007-68034-C03-02 from the Spanish government, 
and FEDER funds. This has been also supported by Plan Andaluz de Investigaci\'on, Desarrollo e Innovaci\'on 
of Junta de Andaluc\'{\i}a as research group FQM322 and excellence fund FQM-5418. J.A.C. is a research 
member of the Consejo Nacional de Investigaciones Cient\'{\i}ficas y Tecnol\'ogicas (CONICET), Argentina. 
The NRAO is a facility of the NSF operated under cooperative agreement by Associated Universities, Inc. 
This publication makes use of data products from the Two Micron All Sky Survey, which is a joint project 
of the University of Massachusetts and the Infrared Processing and Analysis Center/California Institute 
of Technology, funded by the National Aeronautics and Space Administration and the National Science 
Foundation in the USA. This paper is based on observations made with ESO Telescopes at the La Silla 
Paranal Observatory under programme ID $<$71.D-0387$>$. This paper is also based on observations collected 
with the Gran Telescopio de Canarias (GTC) at Observatorio del Roque de los Muchachos (La Palma, Canary 
Islands) operated by the Instituto de Astrof\'{\i}sica de Canarias (IAC). We also thank the anonymous 
referee, whose constructive comments helped us to significatively improve this paper.
}
\end{acknowledgements}

\end{document}